\documentclass[11pt,twoside]{article}
\usepackage{asp2010}

\resetcounters

\begin{document}

\title{Narrow UV Absorption Line Outflows from Quasars}
\author{Fred Hamann,$^1$ Leah Simon,$^2$ Paola Rodriguez Hidalgo,$^3$ and Daniel Capellupo$^1$
\affil{$^1$Department of Astronomy, University of Florida, Gainesville, FL 32611, USA}
\affil{$^2$Department of Physics, Berea College, Berea, KY 40404, USA}
\affil{$^3$Dept. of Astronomy, Penn State University, State College, PA 16801, USA}}

\begin{abstract}
Narrow absorption line (NAL) outflows are an important yet poorly understood part of the quasar outflow phenomenon. We discuss one particular NAL outflow that has high speeds, time variability, and moderate ionizations like typical BAL flows, at an estimated location just $\sim$5 pc from the quasar. It also has a total column density and line widths (internal velocity dispersions) $\sim$100 times smaller than BALs, with no substantial X-ray absorption. We argue that radiative shielding (in the form of an X-ray/warm absorber) is not critical for the outflow acceleration and that the moderate ionizations occur in dense substructures that have an overall small volume filling factor in the flow. We also present new estimates of the overall incidence of quasar outflow lines; e.g., $\sim$43\% of bright quasars have a C~IV NAL outflow while $\sim$68\% have a C~IV outflow line of any variety (NAL, BAL, or mini-BAL). 
\end{abstract}

\section{Introduction}
High velocity quasar outflows appear to be a natural byproduct of accretion onto the 
central super-massive black hole. They appear most conspicuously in quasar 
spectra as broad absorption lines (BALs) with velocity widths $>$2000 km~s$^{-1}$ and maximum speeds that can exceed 0.1$c$ 
\citep{Weymann91}. However, quasar outflows are also detected via narrow absorption lines (NALs,) with FWHM $<$500 km~s$^{-1}$ and often $<$50 km~s$^{-1}$. Recent studies suggest that NAL outflows are even more common than BALs (\S3 below). Thus the NALs are an important piece to the overall puzzle of quasar outflows. They present unique challenges but also unique opportunities to measure the outflow physical properties and constrain theoretical models. In this proceedings, we summarize results from a detailed study of one particular NAL outflow (\S2), and we present new estimates of the incidence of outflow lines in quasar samples (\S3).

\section{High-Velocity Outflow NALs in J2123$-$0050}

The redshift 2.3 quasar J2123$-$0050 has an interesting NAL outflow with five distinct systems spanning velocities $v\sim 9710$ to 14,050 km~s$^{-1}$ and C~IV line widths FWHM $\approx$ 62 to 164 km~s$^{-1}$ \citep[Figure 1,][]{Hamann11}. 
This outflow is remarkable for its high speeds and degree of  
ionization (dominated by C~IV and O~VI) similar to BALs, but line 
widths $\sim$100 times narrower than typical BALs and no significant X-ray 
absorption. The outflow nature of the NALs is confirmed 
by their line variabilities, smooth super-thermal 
line profiles, and doublet ratios that require partial covering of 
the quasar continuum source. The line-of-sight covering fractions derived from doublet ratios indicate characteristic absorber sizes of only 0.01-0.02 pc. 
The line strength variations (on time scales $\leq$0.63 yr in the quasar frame) 
were well coordinated between the five systems and accompanied by nearly commensurate changes in the absorber covering fractions (Figure 2). It seems unlikely that this behavior is caused by the coordinated motions of five distinct outflow structures across our lines of sight. More natural explanations include UV hot spots on the accretion disk and/or global changes in the outflow ionization 
caused by changes in the quasar's ionizing flux. Figure 3 illustrates how changes in the covering fractions can result from changes in the ionization {\it if} the absorber is spatially inhomogeneous. In this figure, the left-hand panel represents a degree of ionization favorable to absorption in a particular ion, such as C~IV. As the ionization changes (moving to the right-hand panels), the ion's column densities and line optical depths decrease, leading to smaller effective covering fractions. 

\articlefigure[scale=0.62]{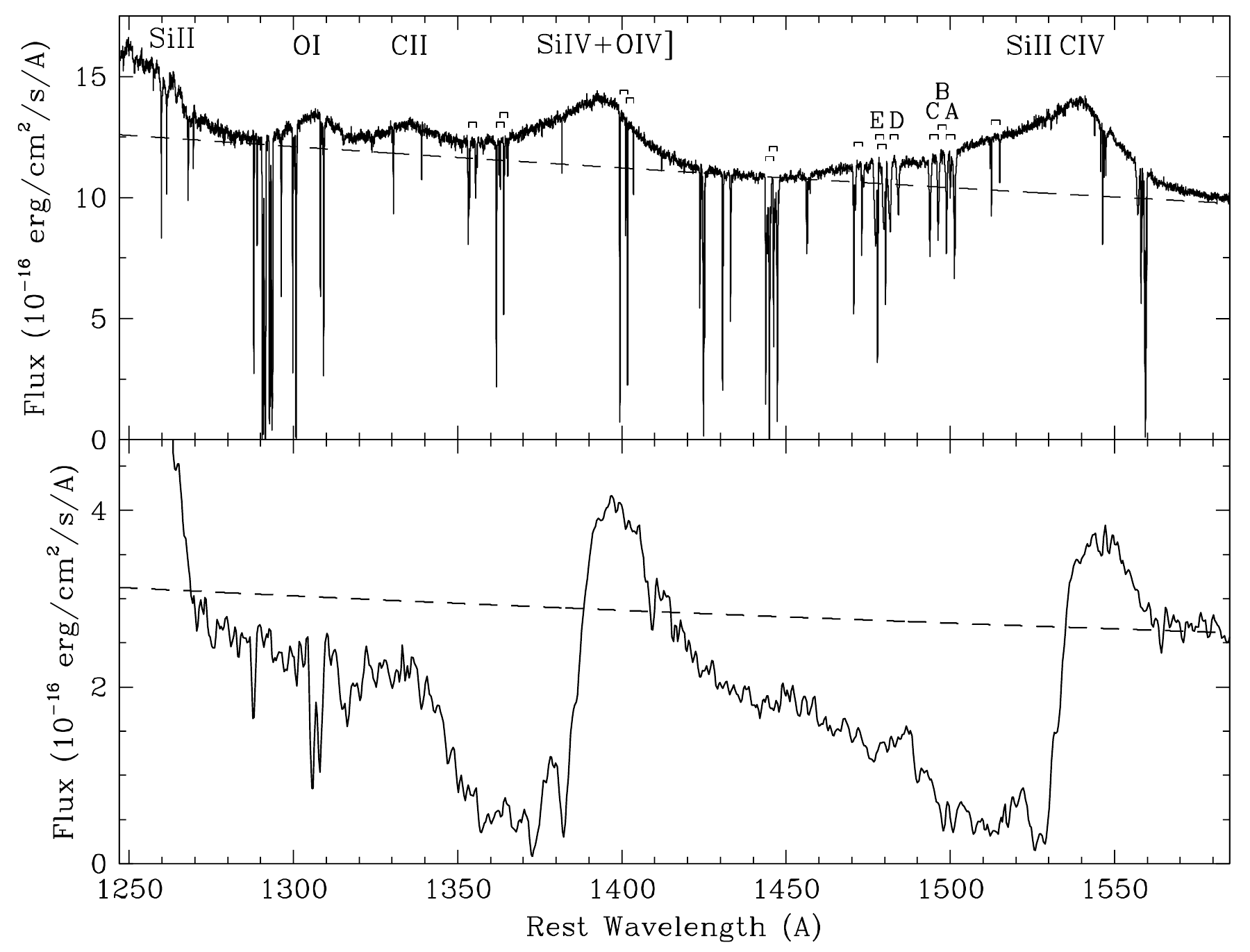}{1}
{{\it Top panel:} Spectrum of the quasar J2123$-$0050 with numerous CIV NAL doublets labeled by open brackets just above the spectrum \citep[from][]{Hamann11}. Most of the NALs probably form in unrelated intervening gas. The confirmed outflow NALs at $v = 9710$ to 14,050 km~s$^{-1}$ are marked by the letters A -- E. The positions of strong broad emission lines are shown across the top. {\it Bottom panel:} Spectrum of the BAL quasar 1331-0108 plotted for comparison to the NAL outflow above. The dashed curves in both panels estimate the continuum-only flux.}

\articlefigure[scale=0.55]{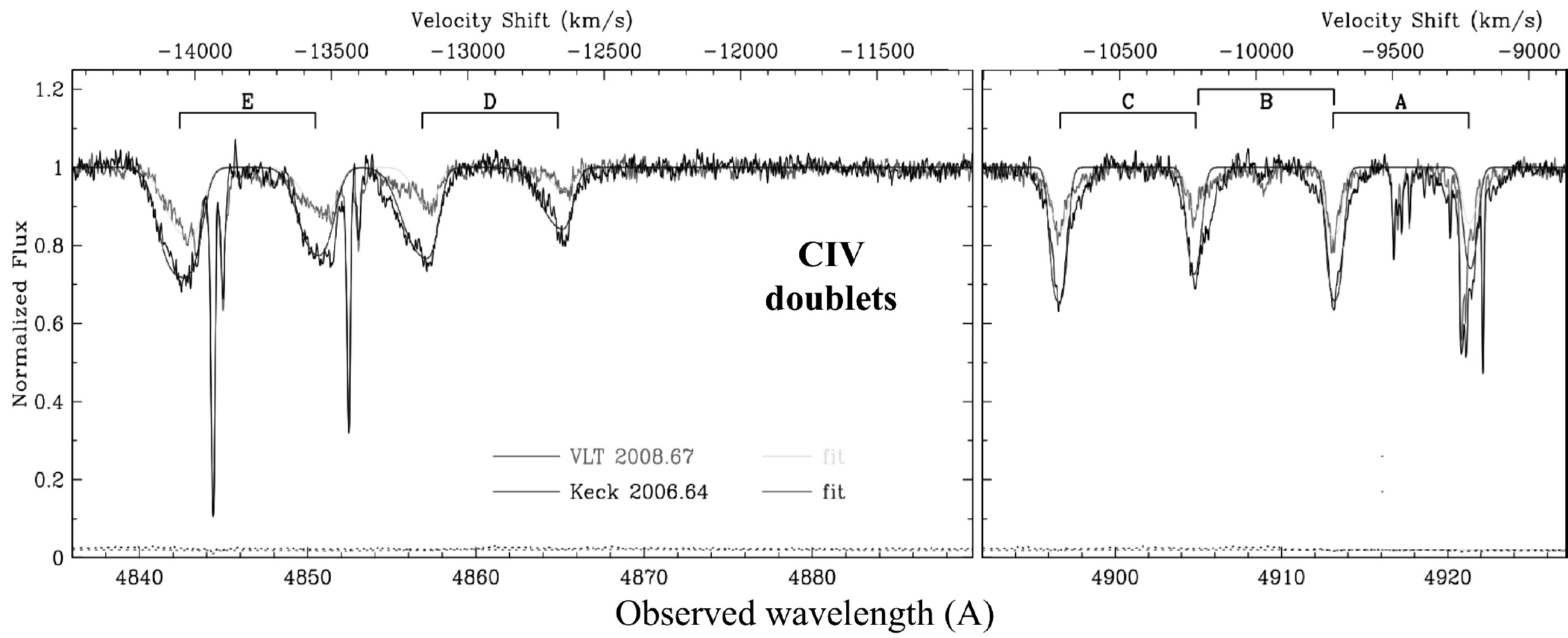}{2}
{Expanded view of the CIV outflow NALs in J2123$-$0050 (cf. Fig. 1) showing their coordinated variabilities between the observed epochs 2006.64 (black spectrum) and 2008.67 (gray with weaker lines). The velocities across the top apply to the shorter wavelength doublet member $\lambda$1548 in the quasar frame.}

\articlefigure[scale=0.54]{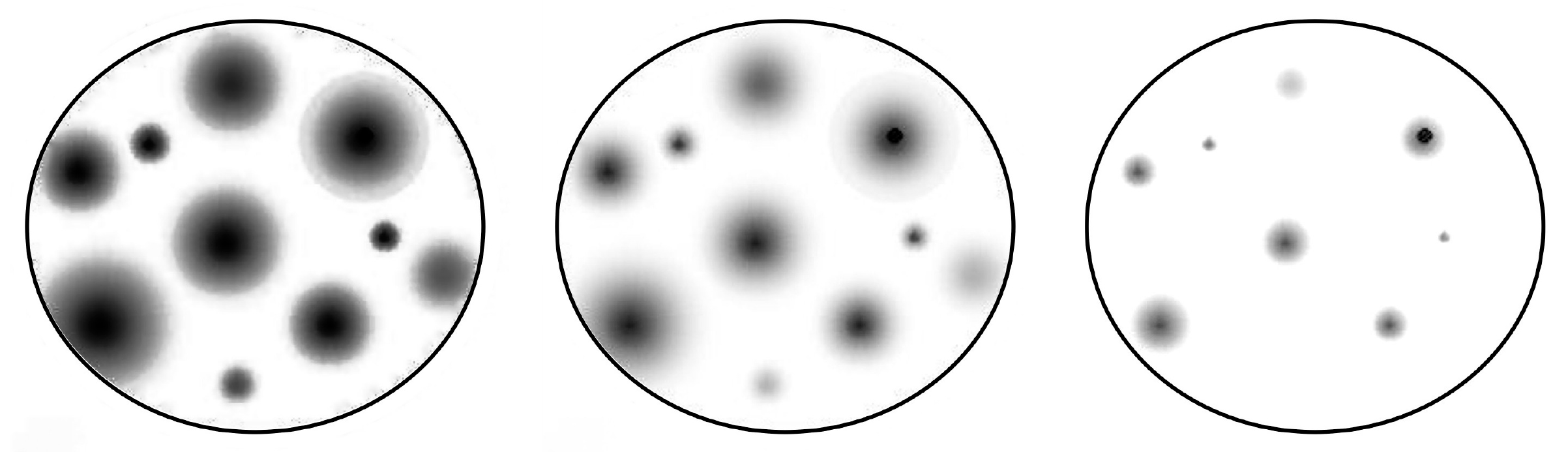}{3}
{Schematic illustrations of inhomogenous partial covering. The background light source (white ovals) is partially obscured by absorbing clouds (gray circles) that each have radially decreasing column densities and line optical depths (indicated by their grayscale opacities). The three panels show how ionization changes in the same absorbing region might change the line optical depths and thus the total effective covering fraction of the clouds. \citep[Adapted from][]{Hamann04}}

Our multi-epoch observations yield an upper limit on the outflow acceleration, $\leq$3 km~s$^{-1}$ yr$^{-1}$, consistent with blobs of gas that are gravitationally unbound and coasting freely $\geq$5 pc from the central black hole. However, these small absorbing structures (based on partial covering) can travel just a few pc before dissipating (without external confinement) and, therefore, the absorbers should be near the $\sim$5 pc minimum radius. The total column density of this flow, $N_H\sim 5\times 10^{19}$ cm$^{-2}$, is at least two orders of magnitude less than the best estimates for BAL outflows, and its kinetic energy luminosity is similarly much too small to affect the host galaxy via feedback. (Note that quasar feedback might still occur in this case if the NAL outflow is accompanied by a much more massive BAL-like flow in other directions, e.g., Fig. 4 below.) 

The absence of strong X-ray absorption is an important result consistent with other NAL and mini-BAL studies \citep{Misawa08,Chartas09,Gibson09a}. It implies that radiative shielding in the far-UV and X-rays is not needed to maintain moderate BAL-like ionizations, and therefore, apparently, it is not needed to facilitate the radiative acceleration to high BAL-like speeds. This contrasts with common ideas about BAL winds, where strong X-ray absorption is usually taken as evidence for an important shielding medium at the base of the flow that prevents the over-ionization and thus facilitates the radiative acceleration of outflow gas farther out \citep{Murray95}. The NAL and mini-BAL data imply that shielding is not essential for the ionization nor, probably, the outflow physics. This conclusion is supported by our recent photoionization calculations (Hamann \& Simon, in prep.) indicating that shielding is, in fact, problematic for the majority of observed quasar outflows because a stationary medium capable of shielding ions like C~IV in the outflow will itself produces strong (narrow) absorption lines at $v\approx 0$ in these same ions. This type of strong $v\approx 0$ line absorption is generally not observed. \cite{Hamann11} argue that the moderate outflow ionizations are achieved, instead, by high gas densities in small sub-structures that occupy an overall small volume filling factor in the flow. 

\section{NALs in the Bigger Picture of Quasar Outflows}

One of the most important questions about quasar NAL outflows is their commonality, e.g., in relation to BALs and mini-BALs. Simon et al. (in prep. and this volume) used partial covering and line profile analyses to identify and measure C~IV outflow NALs in a sample of 24 quasars. They find that $\geq$37\% of quasars have at least one C~IV NAL formed in a quasar-driven outflow at speeds $2500\leq v\leq 40,000$ km~s$^{-1}$. This percentage is a lower limit because the identifications based on partial covering and/or ``broad'' smooth line profiles miss an unknown (but demonstrably significant) number of outflow NALs. However, Simon et al. also note that their quasar sample, which favors sources known previously to have NALs at $v<8000$ km~s$^{-1}$, probably has a weak bias toward more outflow NALs. They balance these factors by adopting $\sim$37\% as their best guess at the true fraction of quasars with a quasar-driven outflow NAL at $v>2500$ km~s$^{-1}$ \citep[also][and Culliton this volume]{Misawa07a}. We can combine this result with \cite{Nestor08}, who measured CIV NALs in a large sample of SDSS quasars. Their analysis of the NAL velocity distribution shows that $\sim$14\% of quasars have at least one quasar-driven C~IV outflow NAL at velocities $v < 12,000$ km~s$^{-1}$ and $\sim$8\% have a quasar-driven outflow NAL at $v < 3000$ km~s$^{-1}$ (see Simon et al. in prep. for more discussion). Once again considering the lower sensitivity and undercounting of outflow NALs in the Nestor et al. study, we use their 8\% to crudely estimate that overall 37\% + 8\% = 45\% of quasars have at least one C~IV outflow NAL. 

This detection rate for NAL outflows is significantly larger than BALs and mini-BALs. In particular, 10-15\% of optically-selected quasars are found to have BALs, although the true fraction is probably closer to $\sim$20\% because quasars with strong BALs are under-represented in quasar surveys \citep{Hewett03,Knigge08}. Similarly, Rodriguez Hidalgo et al. (this volume) find that C~IV mini-BALs appear in $\sim$11\% of SDSS quasars overall and $\sim$5\% of quasars without BALs. Altogether these results imply that roughly 45\% + 20\% + 5\% = 70\% of quasars have a NAL, BAL, or mini-BAL outflow line of C~IV in their spectrum \citep[also][]{Ganguly08}. 

A major challenge now is to understand how NALs, BALs, and mini-BALs fit together into an single paradigm of quasar outflows. Figure 4 shows how these features might coexist in a single outflow geometry \citep[also][]{Elvis00,Ganguly01}. Is this simple unified geometry correct? There surely are strong orientation effects in quasar outflows, probably resembling Figure 4 (e.g., Proga et al. this volume), and it seems natural to attribute the small structures that produce NALs and mini-BALs to the ragged edges of a massive BAL outflow. But evolution might also play a role. The recent work on FeLoBALs \citep[][and this volume]{Farrah10} argues strongly for time-dependent effects, such that stronger/broader lines and larger column densities might be favored generally in younger quasar/accretion disk environments. 

\articlefigure[scale=0.52]{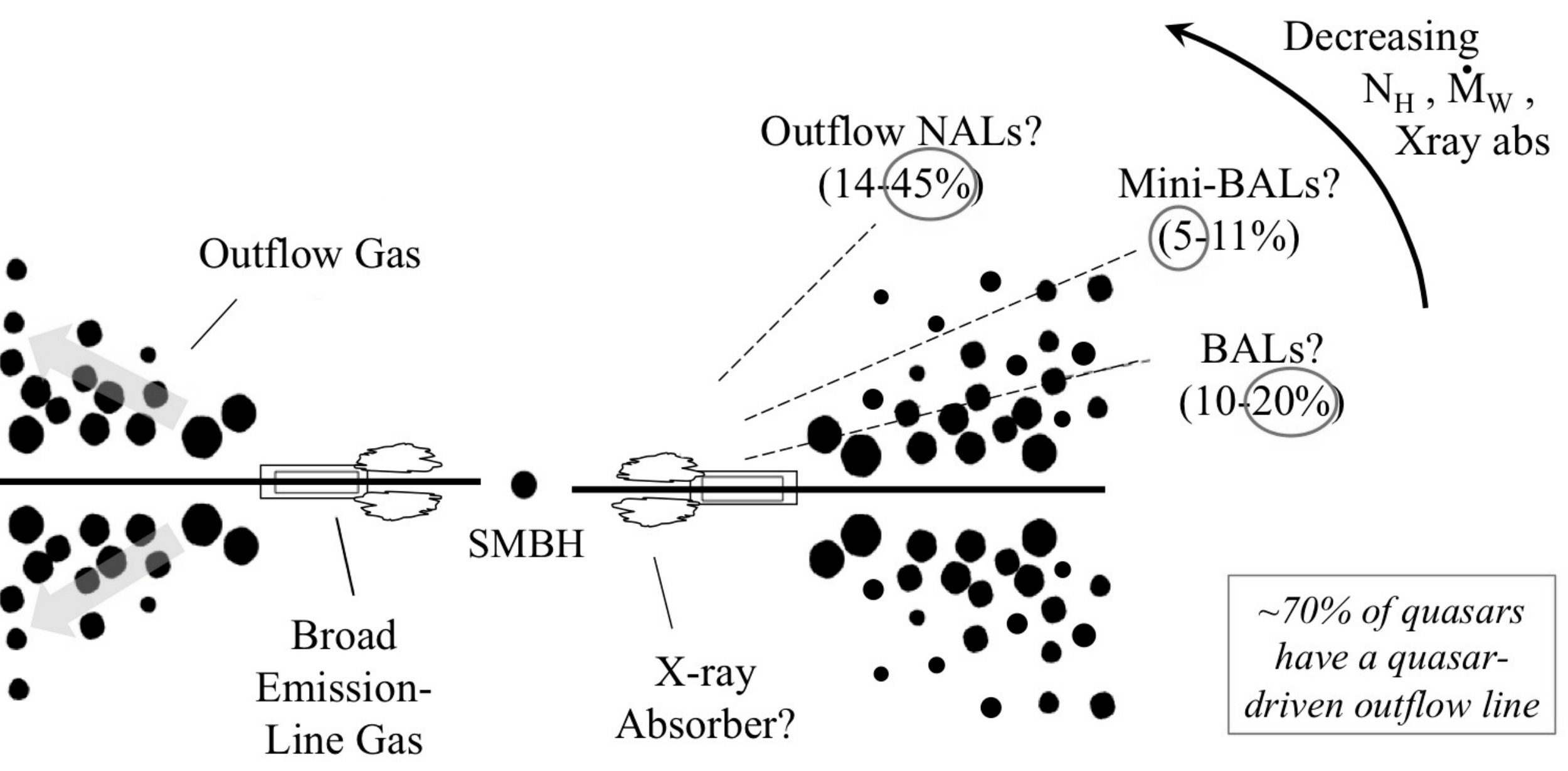}{4}
{A plausible geometry for quasar accretion disk winds. The disk is shown edge-on as a bold horizontal line around the central super-massive black hole (SMBH). The highly ionized X-ray absorber (white clouds above and below the disk) resides somewhere near the base of the outflow (black circles farther out). Different sightlines through this flow (dashed lines) produce BALs, mini-BALs, or outflow NALs in quasar spectra. The trend is for narrower/weaker absorption lines, smaller outflow column densities, and weaker X-ray absorption at higher latitudes above the disk plane. The percent ranges in parentheses are detection frequencies of BALs, mini-BALs, and outflow NALs in quasar spectra (circled values are favored, see text), which might correspond to opening angles for the different outflow types.}

\acknowledgements This work was supported by grants from the National Science Foundation (1009628) the Space Telescope Guest Observer program (GO11705).

\bibliographystyle{asp2010}

\end{document}